\def\rem#1{{\bf #1}}
\def\simless{\mathbin{\lower 3pt\hbox
     {$\rlap{\raise 5pt\hbox{$\char'074$}}\mathchar"7218$}}}   %< or of order
\def\simmore{\mathbin{\lower 3pt\hbox
     {$\rlap{\raise 5pt\hbox{$\char'076$}}\mathchar"7218$}}}   %> or of order
\documentclass[useAMS,usenatbib]{mn2e}
\usepackage{graphicx}

\title[Triple-peaked X-ray burst in 4U 1636--53]{A very rare
triple-peaked type-I X-ray burst in the low-mass X-ray binary 4U
1636--53}

\author[G. Zhang et al.]{Guobao Zhang$^{1}$\thanks{E-mail:
zhang@astro.rug.nl}, Mariano M\'endez$^{1}$, Diego Altamirano$^{2}$,
Tomaso M. Belloni$^{3}$, \newauthor
Jeroen Homan$^{4}$\\
$^{1}$Kapteyn Astronomical Institute, University of Groningen, P.O. BOX
800, 9700 AV Groningen, The Netherlands\\
$^{2}$ Astronomical Institute `Anton Pannekoek', University of
Amsterdam, Science Park 904, 1098 XH, Amsterdam, the Netherlands\\
$^{3}$ INAF -- Osservatorio Astronomico di Brera, Via E. Bianchi 46,
I-23807 Merate (LC), Italy\\
$^{4}$ MIT Kavli Institute for Astrophysics and Space Research, 70
Vassar Street, Cambridge, MA 02139, USA}

\begin{document}

%\pagerange{\pageref{firstpage}--\pageref{lastpage}} \pubyear{2009}

\maketitle

\label{firstpage}

\date{Accepted. Received; in original form}

\begin{abstract}

We have discovered a triple-peaked X-ray burst from the low-mass
X-ray binary (LMXB) 4U 1636--53 with the Rossi X-ray Timing Explorer
(RXTE).  This is the first triple-peaked burst reported from any LMXB
using RXTE, and it is only the second burst of this kind observed
from any source.  (The previous one was also from 4U 1636--53,
and was observed with EXOSAT.) From fits to time-resolved spectra,
we find that this is not a radius-expansion burst, and that the
same triple-peaked pattern seen in the X-ray light curve is also
present in the bolometric light curve of the burst. Similar to what
was previously observed in double-peaked bursts from this source,
the radius of the emitting area increases steadily during the burst,
with short periods in between during which the radius remains more or
less constant. The temperature first increases steeply, and then 
decreases across the burst also showing three peaks. The first and last peak in
the temperature profile occur, respectively, significantly before
and significantly after the first and last peaks in the X-ray and
bolometric light curves. We found no significant
oscillations during this burst. This triple-peaked burst, as well
as the one observed with EXOSAT and the double-peak bursts in this
source, all took place when 4U 1636--53 occupied a relatively narrow
region in the colour-colour diagram, corresponding to a relatively
high (inferred) mass-accretion rate. No model presently available
is able to explain the multiple-peaked bursts.

\end{abstract}


\section{Introduction}

\rem{Thermonuclear, type-I, X-ray bursts (also known as X-ray bursts, or
simply bursts), are due to unstable burning of H and He on the surface
of accreting neutron stars in low-mass X-ray binaries (LMXBs). During
these bursts, the X-ray flux increases by factors of $\sim$10-100 within a  
second or less, after which it returns back to its pre-burst level  
over a time-span of tens of seconds, following an approximately  
exponential decay \citep[e.g.,][]{lewin93, Strohmayer03, galloway}.}

One of the best studied sources of X-ray bursts is the LMXB 4U
1636--53. Also known as V801 Ara, 4U 1636--53 was discovered with OSO-8
\citep{Swank}, and was subsequently studied in great detail using
observations with SAS-3, Hakucho, Tenma, and EXOSAT \citep[see][for a
review]{lewin87}. The orbital period of this binary system is 3.8 hr
\citep{van Paradijs90}, and the spin period of the neutron star is 581
Hz \citep{Strohmayer98a,Strohmayer98b}. Using EXOSAT, \cite{Damen89}
detected 60 bursts from this source between 1983 and 1986; from
observations with the Rossi X-ray Timing Explorer (RXTE), 172 bursts
were detected up to 2007 June 3 \citep{galloway}, and more than 250
bursts including data taken after that date (Zhang et al., in prep.).
Most of these X-ray bursts have standard, single-peaked, fast rising
and exponentially decaying light curves. \rem{Three superbursts have also been detected in this source \citep{wijnands01,Kuulkers04}. \cite{Linares09} detected a 5.4 minutes recurrence time between two X-ray bursts in 4U 1636--53, which is the shortest observed recurrence time in any neutron star low-mass X- ray binary.}

However, there have also been several detections of bursts from 
4U 1636$-$53 that have double-peaked light curves, both with EXOSAT 
\citep{sztajno} and RXTE \citep{Bhattacharyyab, galloway}.
Only one triple-peaked X-ray burst has ever been detected from this 
source by \cite{paradijs} using EXOSAT observations. 
Several models have been proposed to explain the double-peaked bursts.
For instance, \cite{Bhattacharyyab} suggested that these bursts are due
to stalling of the nuclear-burning front near the equator of the
neutron star following ignition near the neutron-star pole, whereas
\cite{fisker} suggested that these bursts are caused by a waiting point
in the thermonuclear reaction chain. 

%\cite{Bhattacharyyab} reported the
%detection of burst oscillations at a $\sim 4\sigma$ level during the
%first peak of a double-peaked burst in 4U 1636--53, at a frequency that
%was consistent with the 581-Hz spin frequency of the neutron star in
%this source \citep{Strohmayer98a,Strohmayer98b}. 

\rem{In this paper}, we report the first detection of a triple-peaked X-ray
burst in 4U 1636$-$53 with RXTE, and only the second ever detected from
this source. The paper is organized as follows: We describe the
observations and data analysis in \S\ref{data}, we show the results in
\S\ref{results}, and finally we discuss our findings in 
\S\ref{discussion}.

\section[]{OBSERVATIONS AND DATA ANALYSIS}
\label{data}

We analysed all data available from the RXTE Proportional Counter Array
(PCA) of 4U 1636--53 as of March 15 2009. The PCA consists of an array
of five collimated proportional counter units (PCUs) operating in the
2$-$60 keV range, with a total effective area of approximately 6500
cm$^2$ and a field of view of $\sim1^\circ$ FWHM \citep{jahoda}. For
each observation we calculated X-ray colours from the Standard2 data
(16-s time-resolution and 129 energy channels in the $2-60$ keV band),
and produced 1-s light curves from the Standard1 data (0.125-s time resolution
with no energy resolution). To calculate the colours of the source, we
first examined the 1-s Standard1 light curves to identify and remove
any X-ray burst from the data, and we subsequently computed light
curves in four energy bands every 16 s from the Standard2 data,
separately for each PCU detector that was active during an observation.
We subtracted the background contribution from each light curve, and we
corrected the count rates for dead time. We defined the soft colour as
the count rate in the 3.5$-$6.0 keV band divided by the count rate in
the 2.0$-$3.5 keV band, and the hard colour as the count rate in the
9.7$-$16.0 keV band divided by the count rate in the 6.0$-$9.7 keV
band. We used data of the Crab pulsar and nebula taken close to each
observation of 4U 1636--53 to correct for instrumental effects
\citep[e.g.][]{Altamirano}. All colours of 4U 1636--53 presented in
this paper are therefore normalised to the colours of Crab. Whenever we
found an X-ray burst, we used 80 s of persistent emission just before
the burst to represent the colours of the source at the time the burst
started. 

In this paper we concentrate on an unusual X-ray burst that took place
at 18:23:25 UTC on December 11 2006 (MJD 54080.76626; ObsID
92023-01-44-10). To calculate the bolometric flux of the persistent
emission before the burst, $F_{\rm pbol}$, we extracted energy spectra
of $\sim 800$ s before the onset of the burst from the two main
instruments on board RXTE, the PCA and the High Energy X-ray Timing
Experiment (HEXTE). HEXTE consists of two clusters of four NaI/CsI
phoswich scintillation detectors that are sensitive to X-rays in the
$15 - 250$ keV range \citep{Gruber96}. Cluster A of HEXTE stopped
working properly in October 2006, an therefore only cluster B was
available for our analysis. For the PCA spectrum we analysed the
Standard2 data of PCU 2, while for the HEXTE spectrum we analysed the
Standard data of cluster-B. We extracted PCA and HEXTE backgrounds, and
produced response matrices for both instruments following the
instructions in the RXTE pages. We added a $0.6$\% systematic error to
the PCU-2 spectrum, but no systematic error was applied to the HEXTE
spectrum. 

To study this burst in detail, we analysed the PCA Event data,
E\_125us\_64M\_0\_1s, in which each individual photon is time tagged at
a $\sim 122$ $\mu$s time resolution in 64 energy channels between
$2-60$ keV. We used all PCUs that were operating at the time of the
X-ray burst (PCUs 0,1,2 and 4) to produce 0.25-s resolution light
curves in the full PCA band and in two other bands, $2.0 - 3.5$ keV and
$6.0 - 9.7$ keV. For the time-resolved spectral analysis of this
burst we extracted spectra in 64 channels every 0.25 s from the Event
data of all available PCUs. Since the light curve of the decay of the
burst is quite smooth, we extracted spectra over somewhat longer
intervals in the tail of the burst to compensate for the lower count
rates. We used a time resolution of 0.5 s, 1 s and 2 s, respectively,
for each of the following time intervals: $20.5-25.5$ s, $25.5-28.5$ s,
and $28.5-32.5$ s after the start of the burst. We generated an
instrument response matrix, and we fitted the spectra using XSPEC
version 12.4.0. Because of the short exposure, in this case the
statistical errors dominate, and therefore we did not add any
systematic error to the spectra. We restricted the spectral fits to the
energy range $3.0 -20.0$ keV. During the $\sim 800$ s previous to the
onset of the burst, the light curve and colours of 4U 1636--53 were
consistent with being constant, therefore we extracted a spectrum of
those data to use as background in our fits; this approach is well
established as a standard procedure in X-ray burst analysis
\citep[e.g.][]{Kuulkers02}. We note that this procedure fails if the
blackbody emission during the burst comes from the same source that
produces the blackbody emission seen in the persistent emission, since
the difference between two blackbody spectra is not a blackbody
\citep{van Paradijs}. This effect is significant only when the net
burst emission is small, and therefore the problems may arise only at
the start and tail of the burst, when the burst emission is comparable
to the persistent emission \cite[see the discussion in][]{Kuulkers02}. We
corrected each spectrum for dead time using the methods supplied by the
RXTE team. 

We computed Fourier power density spectra for the duration of the burst
using the event data over the full PCA band pass. We calculated the
power spectra from 1-s data segments, setting the start time of each
segment to 0.125 s after the start time of the previous segment. Note
that because of this, the individual power spectra are not independent. 

\section[]{RESULTS}
\label{results}

\begin{figure}
\centering
\includegraphics[width=60mm,angle=270]{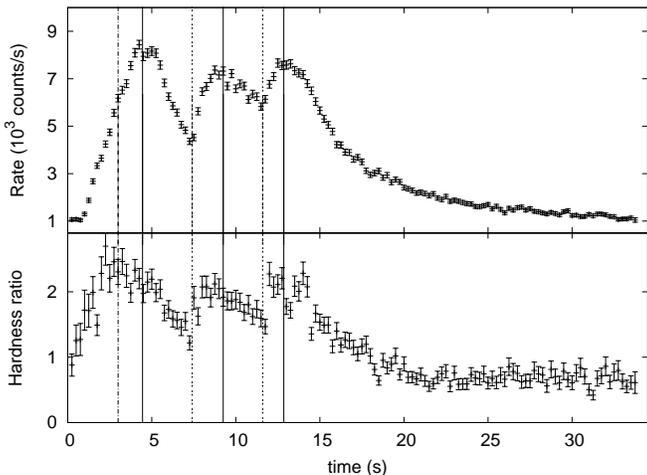}
\caption{Top panel: Light curve in the 2$-$60 keV range of the
triple-peaked burst observed in 4U 1636$-$53. Time is given in seconds
from 2006 December 11 at 18:23:25 UTC. Bottom panel: The time evolution
of the hardness ratio during the triple-peaked burst in 4U 1636$-$53.
The solid vertical lines indicate the times of local maxima in the
light curve, extending into the hardness-curve panel. The dashed
vertical lines indicate the times of local minima in the light curve,
extending into the hardness-curve panel. The left-most dotted-dashed
vertical line indicates the time of the first maximum in the hardness
curve extended into the light-curve panel. We also plotted the $1\sigma$
error bars.} 
\label{lightcurve}
\end{figure}

In the top panel of Figure \ref{lightcurve} we show the light curve of
the burst at a resolution of 0.25 s. The light curve displays 
three peaks, similar to what was observed during the triple-peaked
burst reported by \cite{paradijs} from EXOSAT data of this source, and
we therefore call this a triple-peaked burst. Out of the three peaks in
the burst, the first peak is the brightest in the full PCA band ($\sim
2000$ counts s$^{-1}$ PCU$^{-1}$) and the second is the weakest ($\sim
1700$ counts s$^{-1}$ PCU$^{-1}$). The separation in time between the
first and third peak in the triple-peaked burst reported here is $\sim
8$ s, whereas in the triple-peaked burst in \cite{paradijs} it twice as
long, $\sim 17$ s. In the bottom panel of Figure \ref{lightcurve} we
show the hardness curve during the burst. The hardness is defined as
the count-rate ratio in the $6.0 - 9.7$ keV and $2.0 - 3.5$ keV energy
bands.

Besides the main three peaks, both curves show structure on shorter 
time scales, which makes it difficult to accurately identify the times of the maxima of
each peak. It is nevertheless apparent from this Figure that the first
peak in the light curve occurs significantly later than the first peak
in the hardness curve. Since the hardness is a measure of the
temperature of the emitting surface (see below), this means that the
emitting surface reaches the maximum temperature before the burst is
the brightest. The first peak in the light curve occurs about 2.5 s
after the first peak in the hardness curve. A similar delay between
light curve and hardness curve has been observed in photospheric
radius-expansion bursts \citep{Strohmayer98b}, and is indicative of
the start of the radius-expansion phase but, as we discuss below,
this triple-peaked burst is not a radius expansion burst. After this
initial mismatch between light and hardness curves, the other two
maxima and the two minima of the light curve appear to coincide with
the corresponding maxima and minima in the hardness curve.

\subsection{Pre-burst spectrum}
\label{preburst}

We fitted the PCA (in the $3.0-25.0$ keV range) and HEXTE-B (in the
$15-200$ keV range) pre-burst spectra simultaneously with a model
consisting of a combination of a blackbody, a power law, and a Gaussian
emission line fixed at 6.4 keV, all affected by interstellar
absorption. We included a multiplicative factor in the model to account
for possible systematic differences in the relative calibration of the
effective areas of the two instruments. We found a good fit to the
spectra, $\chi^2 = 92$ for $89$ degrees of freedom (dof), using a model
consisting of a blackbody with temperature $kT = 1.92 \pm 0.05$ keV and
bolometric flux $F_{\rm bb} = 6.3 \pm 0.7 \times 10^{-10} $ erg
cm$^{-2}$ s$^{-1}$, a power law with photon index $\Gamma = 2.96 \pm
0.05$ and normalization $N_{\rm pl} = 1.77 \pm 0.14$ photons cm$^{-2}$
s$^{-1}$ keV$^{-1}$ at 1 keV, and a Gaussian line with $\sigma = 1.08
\pm 0.18$ keV and normalization $N_{\rm G} = 4.2 \pm 1.7 \times
10^{-3}$ photon cm$^{-2}$ s$^{-1}$. \rem{We also included the effect of
interstellar absorption using the cross-sections of \cite{Balucinska92} and solar abundances from \cite{Anders89},} with a hydrogen equivalent column density
fixed at $N_{\rm H} = 0.36 \times 10^{22}$ cm$^{-2}$ \citep{Pandel08}.
These parameters are comparable to those reported by \cite{cackett} for
this source in the soft state. Following \cite{galloway}, we used the
unabsorbed $2.5 - 25$ keV flux, $F_{\rm 2.5-25} = 1.68 \pm 0.02 \times 10^{-9}$
erg cm$^{-2}$ s$^{-1}$, and a bolometric correction $c_{\rm bol} =
1.38$, to estimate the bolometric flux before the burst, \rem{ $F_{\rm pbol}
= 2.32 \pm 0.03 \times 10^{-9}$ erg cm$^{-2}$ s$^{-1}$. We note that the error of 
$c_{\rm bol}$ can be as large as 40 \% \citep{galloway}. Using the
Eddington luminosity for a 1.4-$M_{\odot}$ neutron star with a 10-km
radius, $\gamma = F_{\rm pbol}/F_{\rm Edd} = 3.7 \pm 0.2 \times 10^{-2}$ $-$
$6.2 \pm 0.2 \times 10^{-2}$ for a range
of hydrogen mass fraction $X = 0 - 0.7$.}

\subsection{Time-resolved spectra during the burst}
\label{burst}

We fitted the time-resolved net burst spectra with a single-temperature
blackbody model (bbodyrad in XSPEC), as generally burst spectra are
well fit by a blackbody. During the fitting, we kept the hydrogen
column density $N_{\rm H}$ fixed at $0.36\times10^{22} $cm$^{-2}$
\citep{Pandel08}, and to calculate the radius of the emitting area in
km, we assumed a distance of 5.9 kpc \citep{fiocchi}. The model
provides the blackbody colour temperature ($T_{\rm c}$) and a
normalization proportional to the square of the blackbody radius
($R_{\rm bb}$) of the burst emission surface, and allows us to
estimate the bolometric flux as a function of time.

It is well known \citep{london86,titarchuk94,Madej04} that the colour
temperature $T_{\rm c}$ obtained from fitting the continuum spectra of
an X-ray burst is higher than the effective temperature $T_{\rm eff}$
that enters in the calculation of the bolometric flux. In order to get
the effective temperature $T_{\rm eff}$, we assumed a neutron star with
a mass of $1.4 M_{\odot}$ and a radius of 10 km. Interpolating the
ratio of $T_{\rm c}$/$T_{\rm eff}$ from the table of \cite{Madej04},
and using the bolometric flux from the fit, we calculated the
``corrected'' blackbody radius \citep[see also][]{paradijs}. \rem{We found correction factors between 1.3 and 1.5.}

\begin{figure} 
\centering
\includegraphics[width=100mm,angle=270]{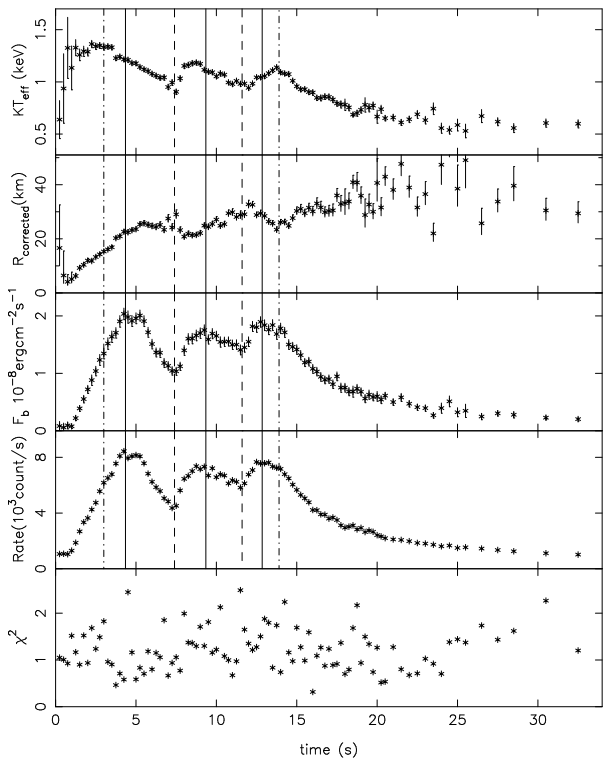}
\caption{Time evolution of spectral parameters during the triple-peaked
burst in 4U 1636--53. From top to bottom the panels show the effective
temperature, the `corrected' (see text for details) radius of the
emitting area, the bolometric flux, the $2-60$ keV light curve, and the
reduced $\chi^2$ of the fits (for 18 dof). The time resolution of this
plot is 0.25 s for times between 0 s and 20.5 s, 0.5 s for times
between 20.5 s and 25.5 s, 1 s for times between 25.5 s and 28.5 s, and
2 s from 28.5 s until the end. The solid vertical lines indicate the
times of local maxima in the X-ray light curve (second panel from the
bottom), extending into the other panels. The dashed vertical lines
indicate the times of local minima in the X-ray light curve extending
into the other panels. The left-most and right-most dotted-dashed
vertical line indicate the times of the first and last maximum in the
temperature curve (upper panel) extended into the other panels. We also
plotted in each panel the $1\sigma$ error bars.}
\label{spectrum}
\end{figure}

In Figure \ref{spectrum} we show the time evolution of the best-fitting
spectral parameters: Effective temperature, $T_{\rm eff}$,``corrected''
blackbody radius, $R_{\rm bb}$, and bolometric flux, $F_{\rm bol}$. For
reference we also plot the $2-60$ keV light curve in this Figure. X-ray
light curves with multiple peaks are often a signature of radius
expansion bursts, since during the radius expansion phase the
temperature of the emitting surface decreases, and part of the
radiation falls outside the X-ray band \citep[e.g.,][]{lewin93}. In
those bursts, however, the bolometric light curve shows a single peak.
In this burst, the bolometric light curve also shows three peaks, at
the same positions as the peaks in the X-ray light curve. This shows
that the multiple peaks in the X-ray light curve are not due to
photospheric radius expansion during the burst, which is consistent
with the fact that the flux at the peak of this burst is about a factor
$2 - 3$ smaller than in real radius expansion bursts in this source
\citep[][]{maurer}. \rem{The total burst fluence was $E_{\rm b} = 2.9 \pm 0.1 \times
10^{-7}$ erg cm$^{-2}$}. From this, and the peak of the bolometric flux,
\rem{the characteristic time scale of the burst was $\tau = E_{\rm b} /
F_{\rm peak} = 13.8 \pm 0.7$ s.} Most of the fits are good, although there are a
few cases with relatively large reduced $\chi^2$ values. (Only in six
occasions the reduced $\chi^2$ is larger than 2.) We note that we did
not add a systematic error to the data (see \S\ref{data}). We confirmed
that the distribution of $\chi^2$ values is consistent with a $\chi^2$
distribution with 18 dof, which is the number of degrees of freedom in
the individual fits.

The top panel of Figure \ref{spectrum} shows that, after an initial
rise, on average the effective temperature decreases with time. The
temperature profile shows three maxima, and as in the case of the
hardness curve, the first peak of the effective temperature takes place
before the first peak of the X-ray and bolometric light curves. The
second peak of the effective temperature profile more or less coincides
with the second peak in the X-ray and bolometric light curves, while
the two minima in the temperature profile also coincide with the times
of minima in the bolometric and the X-ray light curves. The last peak
in the temperature curve occurs significantly after the last peak in
the light curve.

The second panel of Figure \ref{spectrum} shows the evolution of the
radius of the emitting surface, corrected for the fact that the
spectrum during the burst is non-Planckian (see above). On average, the
radius increases steadily during the burst, similar to what was
observed in double-peaked bursts \citep[e.g.][]{Bhattacharyyaa,
Bhattacharyyab}. After initially rising, the radius stops increasing at
the time of the first peak in the light curve, although there is no
apparent change in the trend of the temperature at that point in time.
For $2 - 3$ s the radius remains more or less constant, and after the
second maximum in the light curve, it increases again. More or less at
the time of the second minimum in the light curve the radius stops
increasing, or slightly decreases, and after the third maximum the
radius continues increasing until the end of the burst. (We note that
the slight increase in the radius at the time of minimum temperature
could be due to limitations in the colour correction factor.)

\subsection{Timing analysis of the burst}
\label{timing}

We searched this burst for oscillations around the known spin frequency
of the neutron star in 4U 1636--53 \citep[581 Hz;][]{Strohmayer98a}.
For this, we analysed the data within 15 s of the onset of the burst,
and we searched only the frequency range $579 - 583$ Hz. We chose a
detection limit given by a single trial probability of $3 \times
10^{-6}$ or less of having a given power due to random fluctuations in
the data. This is equivalent to a $\sim4 \sigma$ detection limit
considering the number of trials involved. We did not find any
significant oscillations in this frequency range during the burst. The
95$\%$ confidence upper limits to any oscillation in the frequency
range 579 Hz to 583 Hz were $7.5\%$, $12.3\%$ and $7.5\%$ rms for each
of the three peaks of the burst, respectively, and $12.2\%$ and
$11.3\%$ rms for the rise and the decay of the burst, respectively.
Previously, burst oscillations with amplitudes between $\sim 2\%$ and
$\sim 9\%$ were detected from 4U 1636--53 \citep{zhang w}. 

Most burst oscillations in 4U 1636--53 were so far detected in bright,
photospheric radius expansion bursts \citep{maurer}, with bolometric
peak fluxes larger than $\sim 5.0 \times10^{-8}$ erg s$^{-1}$
cm$^{-2}$. One of the few exceptions to this trend is the possible
detection of oscillations during the first peak of a non-radius
expansion double-peaked burst in \cite{Bhattacharyyab}: The bolometric
flux at the peak in which oscillations were reported was $\sim 1.6
\times10^{-8}$ erg s$^{-1}$ cm$^{-2}$. The bolometric flux at the peak
of the triple-peaked burst reported here was $\sim 2.1 \times10^{-8}$
erg s$^{-1}$ cm$^{-2}$, similar to that of the double-peaked burst of
\cite{Bhattacharyyab}, and a factor $\sim 2-3$ weaker than other
photospheric radius expansion bursts with oscillations in this source.
The upper limit to the rms amplitude we found during the first maximum
of the triple-peaked burst of 4U 1636--53 reported here, $7.5\%$, is
lower than the rms amplitude of the oscillations during the first
maximum of the double-peaked burst, $8.2\%$, reported by
\cite{Bhattacharyyab} from this same source.

\subsection{Colour-colour diagram}
\label{colour}

\begin{figure*}
\centering
\includegraphics[width=100mm, angle=-90]{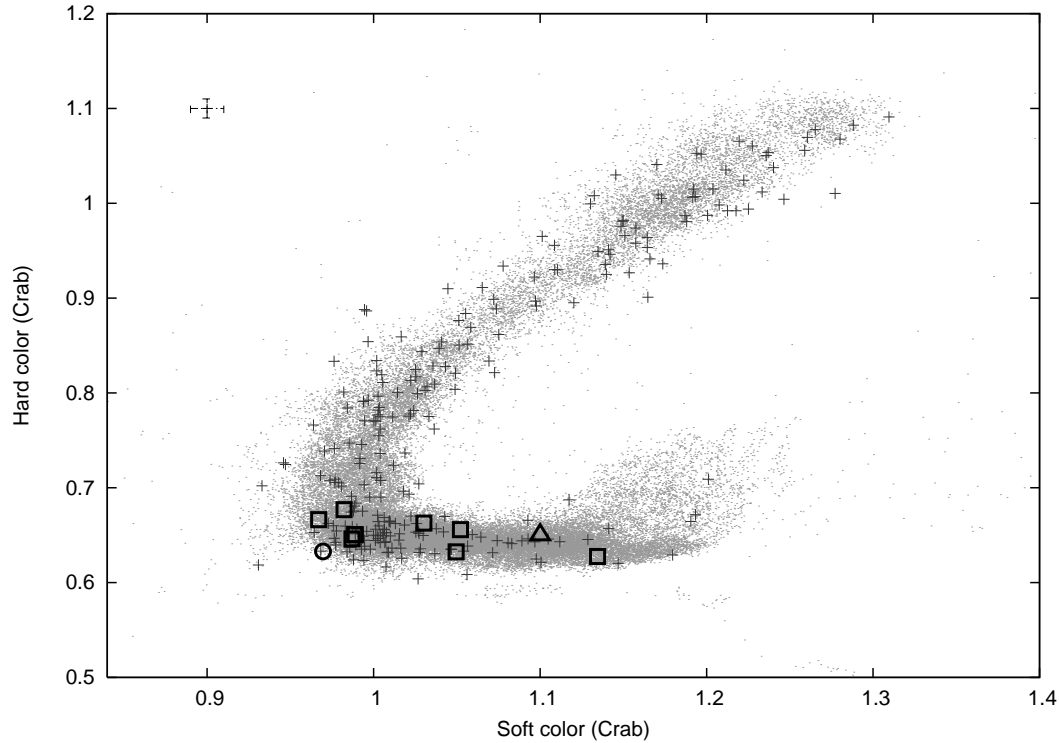}
\caption{Colour-colour diagram of all observations of 4U 1636$-$53 up
to March 2009. The grey points represent the colours of the source from
all RXTE observations available, excluding the times of X-ray bursts
(see text). Each point in this diagram represents 256 s of data. We
plot the representative $1\sigma$ error bar in the upper left part of
this Figure. The black crosses represent the colours of the persistent
emission of the source at the onset of an X-ray burst. The open squares
indicate double-peaked bursts (Galloway et al. 2008; Zhang et al. in
prep.). The open circle indicates the triple-peaked burst. The open
triangle shows the approximate location in this diagram of the
triple-peaked burst observed with EXOSAT \citep{van der Klis90}.} 
\label{ccd}
\end{figure*}

In Figure \ref{ccd} we show the colour-colour diagram of all the RXTE
observations of 4U 1636--53 to date (grey points), excluding times of
X-ray bursts; the black crosses in the plot are the colours of the
source at the start of an X-ray burst. There are more than 250 X-ray
bursts shown in this Figure, from RXTE data taken before March 2009.
The colour-colour diagram of 4U 1636--53 is typical of that of a
low-luminosity LMXB, a so-called Atoll source \citep{Hasinger89}. The
$\sim 250$ X-ray bursts observed with RXTE distribute more or less
uniformly across the colour-colour diagram \citep{Belloni07,Muno04}. 
The triple-peaked burst, which we indicated using a open circle, is
located close to the vertex of this colour-colour diagram, in an area
where there are several other bursts. For comparison, we plot the
approximate position of the triple-peaked burst detected with EXOSAT in
Figure \ref{ccd} \citep{van der Klis90}.

\subsection{Bursts properties}
\label{parameters}

\rem{In order to compare the properties of single-peaked bursts and muti-peaked bursts that occurred in the same area of colour-colour diagram, we selected all the bursts which have hard color less than 0.7 (see Figure \ref{ccd}), and calculated their fluences, peak fluxes and peak temperatures. We found that: (i) Single-peaked bursts have fluences in the range $2 \times 10^{-8}$ to $70 \times 10^{-8}$ ergs cm$^{-2}$, whereas all the multi-peaked bursts have fluences that span a narrower range, from $23 \times 10^{-8}$ to $38 \times 10^{-8}$ ergs cm$^{-2}$; (ii) single-peaked bursts show peak color temperatures from 1.3 to 2.7 keV, while the multi-peaked bursts have temperatures in the range 1.7 to 2.1 keV, with the exception of the double-peaked burst from \cite{Bhattacharyyab}, which has a peak temperature of $2.37 \pm 0.04$ keV. (iii) single-peaked bursts show peak fluxes between 
$3 \times 10^{-9}$ and $82 \times 10^{-9}$ ergs cm$^{-2}$ s$^{-1}$, while the multi-peaked bursts have peak fluxes from $21 \times 10^{-9}$ and $28 \times 10^{-9}$ ergs cm$^{-2}$ s$^{-1}$, except for the same double-peaked burst from \cite{Bhattacharyyab} which has a peak flux of $53 \pm 4 \times 10^{-9}$ ergs cm$^{-2}$ s$^{-1}$ (Zhang et al. in prep.).} 

\section[]{DISCUSSION}
\label{discussion}

We have detected a triple-peaked X-ray burst in the LMXB 4U 1636--53.
This is only the second triple-peaked burst ever detected from any
LMXB, and the first one observed with RXTE. \citep[The only other
triple-peaked burst reported is also from 4U 1636--53, and was observed
more than 20 years ago with EXOSAT;][]{paradijs}. The bolometric light
curve of the triple-peaked burst shows a profile similar to that of the
X-ray light curve, which indicates that this was not a radius expansion
burst (see \S\ref{burst}). This is consistent with the fact that the
bolometric flux at the peak of this burst was a factor of $2-3$ smaller
than the bolometric flux observed from radius-expansion bursts from
this source \cite[e.g.][]{maurer}. 

The temperature curve during this burst shows the same pattern of three
maxima seen in the X-ray light curve, however the first and last peaks
in the temperature curve occur significantly before and significantly
after the first and last peaks in the X-ray light curve, respectively.
The radius of the emitting surface increases more or less steadily
during the burst, with small intervals during which the radius remains
approximately constant or perhaps decreases slightly. 

We found no oscillations in this burst, with upper limits that are
consistent with oscillations detected in previous bursts in 4U
1636--53, but smaller than the amplitude of the oscillations seen
during the non-radius expansion, double-peaked burst in 
\cite{Bhattacharyyab}.

The triple-peaked X-ray burst took place when the source was at an
intermediate position in the colour-colour diagram. There are several
other X-ray bursts around this location in the colour-colour diagram,
none of them showing a triple-peaked profile. If the position of the
source along the $C$-like shape of the colour-colour diagram (Fig.
\ref{ccd}) is a measure of mass accretion rate \citep[with mass
accretion rate increasing as the source moves along the branches from
the upper-right to the lower-left and then to the lower-right corner of
this diagram;][]{Hasinger89}, this indicates that the triple-peaked
nature of this burst is not associated to a rate of mass accretion onto
the neutron star that is different from that in other, single-peaked,
bursts. The previously detected triple-peaked burst in 4U 1636--53 took
place when the source was located in the middle of the branch that
extends to the right of the position of the triple burst in this
diagram \citep{van der Klis90}, when mass accretion rate was presumably
higher. Also in the case of the EXOSAT observations, there are other
normal bursts around the position of triple-peaked burst of
\cite{paradijs} in the colour-colour diagram.

The triple-peaked burst started $\sim 800$ s from the beginning of the
RXTE observation; from the absence of another X-ray burst in those
$\sim 800$ s, the wait time from the previous X-ray burst was $t_{\rm
w} \simmore 800$ s. The previous X-ray burst recorded by RXTE before
the triple-peaked burst took place at 02:42:12 UTC on November 19 2006,
therefore the wait time before the triple-peaked burst was $t_{\rm w}
\simless 2\times10^6$ s. Given the typical burst rate in 4U 1636--53
\citep{galloway}, most likely several X-ray bursts were missed in
between these two observations, \rem{and therefore we can only give a lower
limit of $6.7 \pm 0.1$ to the value of $\alpha = (F_{\rm pbol} t_{\rm w}) /
E_{\rm b}$ of the triple-peaked burst.} 

\cite{boirin} reported X-ray bursts that occur in triplets; we note
that theirs is probably a completely different phenomenon than
the triple-peaked burst that we report here. The bursts detected
by \cite{boirin} are single-peaked, with waiting times of $\sim
12$ min between the three components of the triplet. The burst we
report here shows a triple-peaked structure within a single burst,
with the time between peaks being less than $\sim 10$ seconds.

Most type-I X-ray bursts display a single peak in the bolometric light
curve, with a profile that shows a fast rise and an exponential decay
\citep{galloway}. About a dozen double-peaked and two triple-peaked
X-ray bursts have been detected from the LMXB 4U 1636$-$53
\citep[][this paper]{sztajno, lewin87, paradijs, galloway,
Bhattacharyyaa, Bhattacharyyab, maurer}. Double-peaked bursts were
observed from other sources besides 4U 1636--53, e.g., 4U 1608-52
\citep{Penninx89}, GX 17+2 \citep{Kuulkers02}, and 4U 1709-267
\citep{Jonker04}; but so far there is no report of a triple-peaked
burst from another source besides 4U 1636--52. 

In the past 20 years, several models have been proposed to explain the
double-peaked non-radius expansion bursts. \cite{regev} proposed a
model in which the neutron-star surface is divided in two zones, such
that the burst starts in one zone and moves into the other. In this
model, a temperature gradient develops between the zone where the burst
ignites and the zone onto which the burst flame expands. A heat
transport impedance between the two zones causes a dip in the
temperature and the light curve, and produces a double-peaked burst.
\cite{fisker} repeated these calculations using up two 200 zones, and
found that the effect disappears as the number of zones increases
beyond $\sim 25$. \cite{fujimoto} explained the double-peaked bursts as
a two-step energy generation due to shear instabilities in the fuel on
the stellar surface. \cite{Bhattacharyyaa} argued that this model
cannot reproduce the double-peaked profiles that are observed, and that
it is difficult to maintain sufficient unburnt material in a layer
above the burning flame, as required for the instability to take place,
without having the hot fuel mixing with the cold one. \cite{melia}
suggested that these bursts are due to scattering of the X-ray emission
by material evaporated from the accretion disc during the burst. This
implies that double-peaked bursts can only take place over a certain
range of inclination angles of the binary system. This model, however,
does not explain the evolution of the temperature or the radius of the
emitting area during the bursts, neither does it explain the fact that
only a small fraction of the bursts are multi-peaked. \cite{fisker}
proposed that the double-peaked bursts are due to an interaction
between the shell where the helium flash takes place, and a waiting
point in the $rp$-processes of the shells above. This model can explain
the multi-peaked bursts naturally in terms of a waiting point in the
thermonuclear reaction during the burst, but it is difficult to
reproduce the large dip observed between two peaks in some
double-peaked bursts \citep{Bhattacharyyaa}.

\cite{Bhattacharyyaa, Bhattacharyyab} proposed that the double-peaked
shape of the profile of some type-I X-ray bursts may be due to the
latitude at which the ignition of the burst takes place: For bursts
that ignite at a high latitude, the spreading of the flame on the
surface of the neutron star is opposed by the flow of material accreted
from the disc as this material moves towards the pole. As it approaches
the equator, the burning front stalls momentarily, causing a minimum in
the light curve. Eventually the burning front overcomes the opposing
effect of the accreting material, and extends across the whole surface
of the neutron star, leading to a double-peaked burst. This model can
reproduce both the light curve as well as the spectral evolution of
double-peaked bursts. It also provides an explanation of the
millisecond oscillations detected by \cite{Bhattacharyyab} during the
first peak of a double burst in 4U 1636--53. However, polar ignition (a
crucial factor of this model) should occur at the highest accretion
rates \citep{cooper}, which makes it difficult for this model to
explain that the inferred mass accretion rate at the time of the
double-peaked burst with oscillations \citep{Bhattacharyyab} is lower
than in single-peaked bursts without oscillations \citep{watts}. {\bf Notice, however, that \cite{maurer} suggested that polar ignition can occur in 
a small range of lower accretion rates as well.}
Furthermore, \cite{maurer} simulated burst light-curves for different
ignition latitudes using a phenomenological model for the
time-dependent surface-temperature profile on the neutron-star surface,
and adopting the calculations of \cite{Spitkovsky02} for the speed of
the burning front. 
\cite{maurer} also simulated light curves of bursts igniting quasi-simultaneously in two different places on the neutron-star surface, with delays between ignitions of 0.5 s to 4 s.
None of their simulations produced a multi-peaked profile. 

Our detection of a triple-peaked burst in 4U 1636--53 casts doubt on
the idea that the place of ignition of the burst is the cause of the
multi-peaked light curves \citep{Bhattacharyyaa}. If this scenario
applies to the triple-peaked burst, the burning front must stall twice
before it engulfs the whole surface of the neutron star. In their model
of the double-peaked bursts, \cite{Bhattacharyyaa} argued that the
burning front stalls as it approaches the neutron-star equator, and it
regains speed after it crosses the equator. Clearly, a more complicated
pattern of stalling is required in the case of a triple-peaked burst.

Both double- and triple-peaked bursts in 4U 1636--53 have very
similar properties: (i) The fluence of the triple-peaked burst
reported here is $2.9 \times 10^{-7}$ erg cm$^{-2}$, while the
double-peaked bursts have fluences in the range $2 - 4 \times
10^{-7}$ erg cm$^{-2}$ \cite[see Figure 4 in][]{watts}. (ii)
Except for the double-peaked burst in \cite{Bhattacharyyab}, which
is more or less a factor two brighter than the rest, all double-
and triple-peaked bursts have similar peak-flux values. (iii)
Neither the double- or triple-peaked burst are radius expansion
bursts. (iv) Both triple-peaked bursts \cite[the one reported
here and the previous one reported in][]{paradijs}, and all
double-peaked bursts \citep[][Zhang et al., in prep.]{galloway,
maurer, watts} occur when the source is more or less in the same
region in the colour-colour diagram, accreting at relatively high
inferred mass accretion rate. All these similarities suggest that
the underlying mechanism for double- and triple-peaked bursts should
be the same.  Currently available models are inadequate to explain
the multiple-peaked bursts. It is clear that any model trying to
explain the burst profiles must also be able to explain the fact
that only about a dozen out of $\sim 250$ bursts in the RXTE data of
4U 1636--53 show multiple peaks, with only one out of $250$ in the
RXTE data, and in fact only two out of $\simmore 4.5 \times 10^3$
bursts observed since the beginning of X-ray astronomy \citep{paradijs,
zand, galloway}, showing a triple-peaked profile.

\section*{Acknowledgments} 

It is a pleasure to thank Randall L. Cooper and the referee for helpful
comments. This research has made use of data obtained from the High Energy
Astrophysics Science Archive Research Center (HEASARC), provided by
NASA's Goddard Space Flight Center. TMB acknowledges support from the
Italian Space Agency through grants I/008/07/0 and I/088/06/0. JH
is grateful for support from NASA, through grants NNG05G053G and
NNX06AE22G

\label{lastpage}

\end{document}